\documentclass[12pt]{article}%
\usepackage{amsmath,latexsym}
\usepackage{graphicx}
\usepackage{amsmath}
\usepackage{amsfonts}
\usepackage{amssymb}%
\begin{document}

\title{{Seeking connections between wormholes, gravastars, and
  black holes via noncommutative geometry}}
   \author{
Peter K. F. Kuhfittig* and Vance D. Gladney*
\\  \footnote{kuhfitti@msoe.edu}
 \small Department of Mathematics, Milwaukee School of
Engineering,\\
\small Milwaukee, Wisconsin 53202-3109, USA}

\date{}
 \maketitle

\begin{abstract}\noindent
Noncommutative geometry, an offshoot of string
theory, replaces point-like objects by smeared
objects.  The resulting uncertainty may cause
a black hole to be observationally
indistinguishable from a traversable wormhole,
while the latter, in turn, may become
observationally indistinguishable from a
gravastar.  The same noncommutative-geometry
background allows the theoretical construction
of thin-shell wormholes from gravastars and
may even serve as a model for dark energy.\\

\\
\\
\noindent
PAC numbers: 04.20.-q, 04.20.Jb
\end{abstract}

\section{Introduction}\label{E:introduction}

The purpose of this paper is to examine the
effect of noncommutative geometry on several
diverse phenomena.  These include traversable
wormholes, gravastars, and thin-shell
wormholes from gravastars.  The
noncommutative-geometry background itself
may even serve as a model for dark energy.

Noncommutative geometry, an offshoot of string
theory, replaces point-like objects by smeared
objects \cite {SS03, NSS06, NS10}.  As a result,
spacetime can be encoded in the commutator
 $[\textbf{x}^{\mu},\textbf{x}^{\nu}]
=i\theta^{\mu\nu}$, where $\theta^{\mu\nu}$ is
an antisymmetric matrix that determines the
fundamental cell discretization of spacetime
in the same way that Planck's constant
discretizes phase space \cite{NSS06}.  An
efficient way to model the smearing effect,
discussed in Refs. \cite{LL12, NM08, pK15},
is to assume that the energy density of the
static and spherically symmetric and
particle-like gravitational source is given
by
\begin{equation}\label{E:rho}
  \rho(r)=\frac{\mu\sqrt{\beta}}
     {\pi^2(r^2+\beta)^2},
\end{equation}
which can be interpreted to mean that the
gravitational source causes the mass $\mu$
of a particle to be diffused throughout the
region of linear dimension $\sqrt{\beta}$
due to the uncertainty; $\sqrt{\beta}$ has
units of length.  Eq. (\ref{E:rho}) leads
to the mass distribution
\begin{equation}\label{E:source}
   \int^r_04\pi(r')^2\rho(r')dr'=
   \frac{2M}{\pi}\left(\text{tan}^{-1}
   \frac{r}{\sqrt{\beta}}-
   \frac{r\sqrt{\beta}}{r^2+\beta}\right),
\end{equation}
where $M$ is now the total mass of the source.

According to Ref. \cite{NSS06}, noncommutative
geometry is an intrinsic property of spacetime
and does not depend on any particular features
such as curvature and can therefore be
extremely useful.  This usefulness is further
emphasized in Ref. \cite{NSS06}: there seems
to be a need to modify the four-dimensional
Einstein action to incorporate the noncommutative
effects.  It is shown, however, that the effects
of noncommutativity can be taken into account by
keeping the standard form of the Einstein tensor
on the left-hand side of the field equations and
inserting the modified energy-momentum tensor as
a source on the right-hand side.  This leads to
the conclusion, also discussed in Ref. \cite{RJ16},
that the length scales can be macroscopic.  This
scale allows the determination of the mass
distribution in Eq. (\ref{E:source}), as well as
the application to macroscopic objects such as
wormholes.

To illustrate an important aspect of
nomcommutative geometry, let us start with
the Schwarzschild line element
\begin{equation}\label{E:line1}
ds^{2}=-\left(1-\frac{2M}{r}\right)dt^{2}+
\frac{dr^2}{1-2M/r}+r^{2}(d\theta^{2}
+\text{sin}^{2}\theta\,d\phi^{2}).
\end{equation}
Now denote the right-hand side of Eq.
(\ref{E:source}) by $M_{\beta}(r)$.
Used in Eq. (\ref{E:line1}), we get
\begin{equation}\label{E:regular}
ds^{2}=-\left(1-\frac{2M_{\beta}(r)}{r}
\right)dt^{2}+\frac{dr^2}
{1-2M_{\beta}(r)/r}+r^{2}(d\theta^{2}
+\text{sin}^{2}\theta\,d\phi^{2}).
\end{equation}
By letting $\beta\rightarrow 0$, we
recover the Schwarzschild line element.
However, since
\[
   \text{lim}_{r\rightarrow 0}
   \frac{M_{\beta}(r)}{r}=0,
\]
there is no curvature singularity, a
consequence of having replaced the
point-like object at $r=0$ by a
smeared object.  The result is a
regular black hole.

Noncommutative geometry is by no means
the only modified gravitational theory
to have had a profound impact on the
above topics.  For example, on the
subject of wormholes, Lobo and Oliveira
\cite{LO09} investigated such structures
in the context of $f(R)$ modified gravity,
while noncommutative-geometry inspired
wormholes in $f(R)$ gravity are
analyzed in Ref. \cite{pK18}.  Modified
gravitational theories also play a key
role in the study of compact stellar
objects, including gravastars.  In
particular, the effect of an
electromagnetic field on isotropic
gravastar models in $f(R,T)$ gravity
is discussed by Z. Yousaf; included
among its features is a stability
analysis.  (See Ref. \cite{zY1} and
references therein.)  Moreover,
self-gravitating spherical stars
evolving in the presence of an imperfect
fluid in $f(R,T)$ gravity may exhibit
an irregular energy density whose causes
are investigated in Ref. \cite{zY2}.
The influence of the same modified
gravitational theory on the dynamics
of radiating spherical fluids is
presented in Ref. \cite{zY3}.  Other
noteworthy references involving $f(G,T)$
modified gravity are Refs. \cite{
zY4, zY5}.

Wormhole solutions in $f(T)$ modified
gravity, again with a
noncommutative-geometry background, are
studied in Ref. \cite{mS13}, while
galactic-halo wormholes in $f(T)$
gravity are discussed in Ref.
\cite{mS14}.

\section{The approximate metric}

Consider the line element
\begin{equation}\label{E:line2}
ds^{2}=-\left(1-\frac{2M}{r}\right)dt^{2}+
\frac{dr^2}{1-m(r)/r}+r^{2}(d\theta^{2}
+\text{sin}^{2}\theta\,d\phi^{2}).
\end{equation}
To accommodate this metric, the effective
mass $m(r)$, again determined by integration,
must have the form
\begin{equation}\label{E:m(r)}
   m(r)=\frac{4M}{\pi}\left(\text{tan}^{-1}
   \frac{r}{\sqrt{\beta}}-
   \frac{r\sqrt{\beta}}{r^2+\beta}\right)
   +C_{\beta},
\end{equation}
where $C_{\beta}>0$ is a set of arbitrary
constants such that
$\text{lim}_{\beta\rightarrow 0}C_{\beta}=0$.
The reason is that
\begin{equation}
    \text{lim}_{\beta\rightarrow 0}m(r)=2M,
\end{equation}
thereby recovering the Schwarzschild line
element. The meaning of $m(r)$ in Eq.
(\ref{E:line2}) now becomes clear: it is
a smeared mass that, seen from a distance,
is simply $2M$.  So line element (\ref{E:line2})
approximates the Schwarzschild metric
(\ref{E:line1}).

Before continuing, let us list the Einstein
field equations using the more general metric
from Ref. \cite{MWT}:
\begin{equation}
ds^{2}=-e^{2\Phi(r)}dt^{2}+\frac{dr^2}{1-m(r)/r}
+r^{2}(d\theta^{2}+\text{sin}^{2}\theta\,
d\phi^{2}).
\end{equation}
The equations are
\begin{equation}\label{E:Einstein1}
  \rho(r)=\frac{m'}{8\pi r^2},
\end{equation}
\begin{equation}\label{E:Einstein2}
   p_r(r)=\frac{1}{8\pi}\left[-\frac{m}{r^3}+
   2\left(1-\frac{m}{r}\right)\frac{\Phi'}{r}
   \right],
\end{equation}
\begin{equation}\label{E:Einstein3}
   p_t(r)=\frac{1}{8\pi}\left(1-\frac{m}{r}\right)
   \left[\Phi''-\frac{m'r-m}{2r(r-m)}\Phi'
   +(\Phi')^2+\frac{\Phi'}{r}-
   \frac{m'r-m}{2r^2(r-m)}\right].
\end{equation}
So from Eq. (\ref{E:Einstein1}) and Eq.
(\ref{E:rho}), it follows that
\begin{equation}
   0<m'(r)<1,
\end{equation}
since $\rho >0$.

\subsection{Wormholes}\label{S:wormhole}
Wormholes are handles or tunnels in spacetime
that are able to link widely separated regions
of our Universe or entirely different universes.
Morris and Thorne \cite{MT88} proposed the
following static and spherically symmetric
line element for the wormhole spacetime:
\begin{equation}\label{E:line3}
ds^{2}=-e^{2\Phi(r)}dt^{2}+\frac{dr^2}{1-b(r)/r}
+r^{2}(d\theta^{2}+\text{sin}^{2}\theta\,
d\phi^{2}),
\end{equation}
using units in which $c=G=1$.  Here $b=b(r)$
is called the \emph{shape function} and
$\Phi=\Phi(r)$ is called the \emph{redshift
function}, which must be everywhere finite
to avoid an event horizon.  For the shape
function we must have $b(r_{\text{th}})=
r_{\text{th}}$, where $r=r_{\text{th}}$ is
the radius of the \emph{throat} of the wormhole.
An important requirement is the
\emph{flare-out condition} at the throat:
$b'(r_{\text{th}})<1$, while $b(r)<r$ near the
throat.  The flare-out condition can only
be met by violating the null energy
condition, which states that for the
energy-momentum tensor $T_{\alpha\beta}$,
$T_{\alpha\beta}k^{\alpha}k^{\beta}\ge 0$
for all null vectors $k^{\alpha}$.  (Here
$T^t_{\phantom{tt}t}=-\rho$ is the energy
density, $T^r_{\phantom{rr}r}= p_r$ the
radial pressure, and
$T^\theta_{\phantom{\theta\theta}\theta}=
T^\phi_{\phantom{\phi\phi}\phi}=p_\bot$
the lateral pressure.)  For example, if the
outgoing null vector
$(1,1,0,0)$ yields $\rho+p_r<0$, the condition
is violated.  Matter that violates the null
energy condition is called ``exotic" in Ref.
\cite{MT88}.

Instead of a throat, the Schwarzschild black
hole has an event horizon at $r=2M$, making it
a very different structure.  On the other hand,
their topological similarities have suggested a
deeper connection: according to Hayward \cite
{sH02}, if enough exotic matter is pumped into
a black hole, it becomes a traversable wormhole.
In this paper, we will use noncommutative
geometry to show that the characteristic
uncertainty could make the black hole
observationally indistinguishable from a
wormhole.  Such a structure is sometimes
called a black-hole mimicker
\cite{DS07, LZ07, LZ08, BZ15}.

Since the function
\begin{equation}\label{E:f(r)}
   f(r)=\frac{4M}{\pi}\left(\text{tan}^{-1}
   \frac{r}{\sqrt{\beta}}-
   \frac{r\sqrt{\beta}}{r^2+\beta}\right)
\end{equation}
approaches $2M$ from below, $C_{\beta}$ needs
to be large enough so that $m(r)$ has the form
in Fig. 1.  Thus $m(r)$ is a translation of
\begin{figure}[tbp]
\begin{center}
\includegraphics[width=0.8\textwidth]{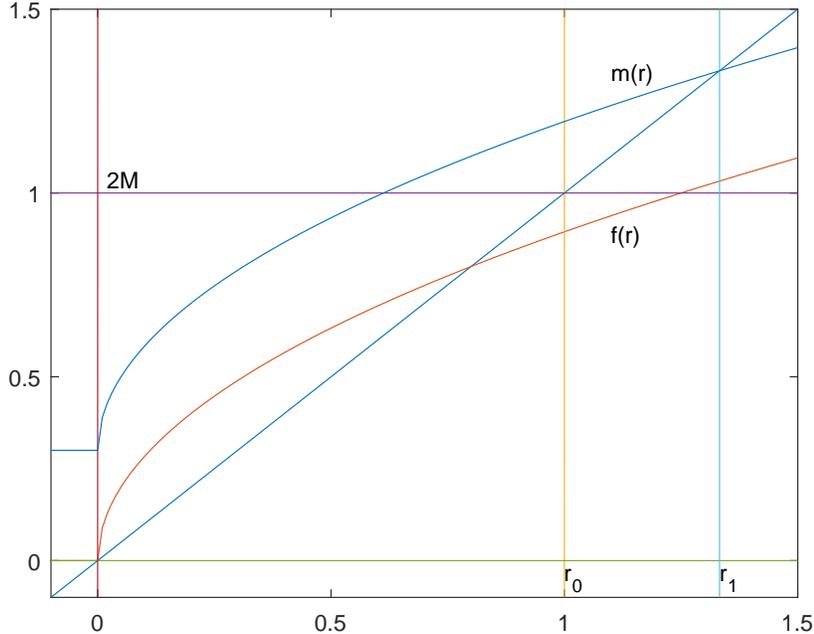}
\end{center}
\caption{$r=r_0$ is the radius of the event horizon
   and $r=r_1$ is the radius of the throat.}
\end{figure}
$f(r)$ in the vertical direction.  Here $r=r_0$
is the radius of the event horizon.  Since the
slope of $m(r)$ is strictly less than unity,
$m(r)$ crosses the 45-degree line at some
$r=r_1$.  As a result,
\[
  m(r_1)=r_1\quad\text{and}\quad 0<m'(r_1)<1,
\]
$r_1$ being the radius of the throat, thereby
satisfying all the requirements of the
shape function of a Morris-Thorne wormhole.
For $r\ge r_1$, the redshift function
$\Phi=\Phi(r)$ can be determined from
$e^{2\Phi(r)}=1-2M/r$, i.e.,
$\Phi(r)=\frac{1}{2}\,\text{ln}\,(1-2M/r)$,
leading to
\begin{equation}
   \Phi'(r_1)=\frac{M}{r_1^2}\frac{1}{1-2M/r_1}.
\end{equation}
So if $r_1$ is sufficiently close to $r_0$, the
tidal forces become so large that the wormhole
becomes observationally indistinguishable
from a black hole, at least in the short run.
(Eventually, the Hawking radiation would have
an observable effect.)

According to Ref. \cite{NSS06}, there is no
need to change the Einstein tensor in the
field equations since the noncommutative
effects can be implemented by modifying
only the stress-energy tensor.  So the
length scales can be macroscopic.  As a
result, the constant $C_{\beta}$ can be
large and still lead to a valid solution.
To be consistent with observation, however,
$C_{\beta}$ is likely to be confined to a
narrow range with $r_1$ close to $r_0$,
making the wormhole a black-hole mimicker.

An alternative interpretation, also based
on noncommutative geometry, is discussed
in the next subsection.

\subsection{Gravastars}\label{SS:gravastar}

Returning to Subsection \ref{S:wormhole}, a
traversable wormhole results from an
appropriate choice of $C_{\beta}$ within a
narrow range and with $r_1$ close to $r_0$.
The noncommutative-geometry background
suggests another interpretation that
remains valid for larger $r$.  First we
observe that decreasing $\beta$ in Eq.
(\ref{E:f(r)}) has the same effect as
increasing $r$, i.e.,
$\text{lim}_{\beta\rightarrow 0 }f(r)=
\text{lim}_{r\rightarrow \infty}f(r)
=2M$.  So for sufficiently small $\beta$,
$f(r)$ will converge to $2M$ so rapidly
that $f(r)$ is approximately constant (thus
$m'(r_1)\approx 0$).  So while this may not
exclude a wormhole with a throat at $r=r_1$,
the structure would become
observationally indistinguishable from an
ordinary spherical shell of thickness
$\delta$, which could be described as
$\delta=r_2-r_1$ for some $r_2>r_1$, and
becoming  extremely dense whenever $M$
is sufficiently large.  Since such a shell
would consist of ordinary matter, its
equation of state becomes $p(r)=\rho(r)$.
(For the exterior region $r>r_2$, we have
$p(r)=\rho(r)=0$.)  Such an extremely dense
shell would ordinarily collapse, but, given
our noncommutative-geometry background, we
can return to Ref. \cite{NSS06} to note that
the relationship between the radial pressure
and energy density is given by
\begin{equation}\label{E:EOS}
   p_r(r)=-\rho(r).
\end{equation}
The reason is that the source is a
self-gravitating droplet of anisotropic fluid
of density $\rho$, and the radial pressure
is needed to prevent a collapse to a matter
point.  So far, then, the properties of the
gravastar are correlated with our
noncommutative-geometry background.
However, having been reduced to an ordinary
spherical volume, we would expect the interior
pressure to be isotropic.  Indeed, from Eq.
(\ref{E:rho}), we have
\begin{equation}\label{E:radial}
   p_r=-\rho=-\frac{\mu\sqrt{\beta}}
   {\pi^2(r^2+\beta)^2.};
\end{equation}
now, according to Ref. \cite{NSS06}, the
lateral pressure $p_{\bot}$ is given by
\begin{equation}
   p_{\bot}=-\rho-\frac{r}{2}\frac
   {\partial\rho}{\partial r}
   =-\frac{\mu\sqrt{\beta}}
   {\pi^2(r^2+\beta)^2}
   +\frac{2\mu r^2\sqrt{\beta}}
   {\pi^2(r^2+\beta)^3}.
\end{equation}
For larger $r$, this reduces to
\begin{equation}\label{E:lateral}
   p_{\bot}=-\frac{\mu\sqrt{\beta}}
   {\pi^2(r^2+\beta)^2}
\end{equation}
in agreement with Eq. (\ref{E:radial}).
Thanks to the noncommutative-geometry
background, we can therefore take the
equation of state for the interior to be
\begin{equation}\label{E:isotropic}
   p(r)=-\rho(r).
\end{equation}
Eq. (\ref{E:isotropic}) makes sense in another
way: it describes a de Sitter vacuum that
counteracts the extreme tension from the
dense shell.

Taken together, these results can serve as
a model of a gravastar, as described in
Refs. \cite{MM02, MM04}.  In other words,
the interior volume contains an isotropic de
Sitter vacuum, while the exterior is
characterized by a Schwarzschild geometry,
separated by a thin shell of stiff matter,
i.e., there are three regions with the
following respective equations of state:\\
\phantom{aa} 1.  interior: $0\le r<r_1$,\quad
      $p=-\rho$;\\
\phantom{aa} 2. shell: $r_1<r<r_2$, \quad
    $p=+\rho$;\\
\phantom{aa} 3. exterior: $r>r_2$, \quad
      $p=\rho=0$.\\
We conclude that a noncommutative-geometry
wormhole may be observationally
indistinguishable from a gravastar and
could therefore be called a
gravastar-mimicker.

\emph{Remark:} While the properties of
the gravastar can be accounted for in
the context of noncommutative geometry,
these properties have analogues in other
theoretical approaches.  An example is
the electromagnetic field discussed in
Ref. \cite{zY1}.  More precisely, just
as the electromagnetic charge affects
the pressure-density profile, the
proper length of the thin shell and
its energy content, the
noncommutative-geometry background
produces the interior isotropic
pressure, as well as an extremely
dense shell.

\section{Additional aspects}
So far we have concentrated on certain
observational effects of a
noncommutative-geometry background
caused by the uncertainty, to which we
may now add another, somewhat simpler,
wormhole model.  Two other aspects,
also discussed in this section, are
concerned with thin-shell wormholes from
gravastars, as well as a possible model
for dark energy.

\subsection{A special wormhole}
Consider the metric
\begin{equation}\label{E:line4}
ds^{2}=-\left(1-\frac{f(r)}{r}\right)dt^{2}+
\frac{dr^2}{1-2M/r}+r^{2}(d\theta^{2}
+\text{sin}^{2}\theta\,d\phi^{2}),
\end{equation}
where $f(r)$ is defined in Eq.
(\ref{E:f(r)}).  As before, if
$\beta\rightarrow 0$, we recover the
Schwarzschild line element.  Now recall
from Fig. 1 that for $\beta >0$, $f(r)<2M$
in the vicinity of $r=r_0$.  In particular,
$f(r_0)<2M$.  It follows that Eq.
(\ref{E:line4}) represents a wormhole
with $b(r)=2M$; $r=r_0$ is the radius
of the  throat.  Since $f(r_0)$ is
close to $2M$,  this wormhole is also
a black-hole mimicker.

\subsection{Thin-shell wormholes from
   gravastars}
In this section we will consider the
theoretical construction of thin-shell
wormholes using the standard cut-and-paste
technique originally due to Visser
\cite{PV95}.  The difference is that
instead of a Schwarzschild black hole,
we employ a gravastar combined with a
noncommutative-geometry background.
Accordingly, following Ref. \cite{PV95},
we start with two copies of a gravastar
and remove from the interior of each
the four-dimensional region
\begin{equation}\label{E:remove}
  \Omega^\pm = \{r\leq a\,|\,a<r_g\},
\end{equation}
where $r=r_g$ is the inside radius of the
stiff-matter shell of the gravastar.  Now
identify the time-like hypersurfaces
\begin{equation}
  \partial\Omega^\pm =\{r=a\,|\,a<r_g\}.
\end{equation}
The resulting manifold is geodesically
complete and possesses two regions
connected by a throat.

Since the spherical surface $r=a$ is
infinitely thin, $p_r=0$.  From the
equation of state $p_r=-\rho$, we also
have $\rho =0$.  So for the null
vector $(1,0,0,1)$, we have from
Eq. (\ref{E:lateral}),
\begin{equation}
   \rho +p_{\bot}
   =0-\frac{\mu\sqrt{\beta}}
   {\pi^2(r^2+\beta)^2}<0,
\end{equation}
so that the null energy condition is
violated, thereby fulfilling a key
requirement for maintaining a
wormhole.

The question is whether such a wormhole
can be stable.  The de Sitter vacuum
inside the gravastar prevents a collapse
of the thin shell.  Normally, however,
the same de Sitter vacuum would cause
the thin shell to explode, but this
outcome is prevented by the stiff
shell of the gravastar.

By contrast, thin-shell wormholes from
Schwarzschild black holes are unstable
unless the velocity of sound exceeds
the velocity of light.

\emph{Remark:} Thin-shell wormholes
from compact stellar objects are
discussed in Ref. \cite{pK17}.

\subsection{Dark energy}
Up to now our primary concern has been
wormholes and gravastars, but our
noncommutative geometry background can
also be applied to a more general
cosmological setting.  Here we would be
dealing with a perfect fluid with the
barotropic equation of state $p=
\omega\rho$, where $\omega$ is a constant.
Now, by Eq. (\ref{E:isotropic}), $\omega =
-1$, which corresponds to Einstein's
cosmological constant, still considered
to be the best model for dark energy
\cite{rB}.  The reason is that dark
energy is vacuum energy with the
observed cosmological constant
$(1.35\pm 0.15)\times 10^{-123}$ in
Planck units.

These observations are consistent with the
accelerated expansion $\overset{..}{a}(t)
>0$, referring to the Friedmann equation
\begin{equation}\label{E:Friedmann}
   \frac{\overset{..}{a}(t)}{a(t)}
   =-\frac{4\pi}{3}(\rho+3p).
\end{equation}
By Eq. (\ref{E:isotropic}), $p=-\rho$;
so it follows directly that
\[
   -\frac{4\pi}{3}(\rho-3\rho)>0.
\]

\section{Conclusion}

Noncommutative geometry, an offshoot of
string theory, replaces point-like objects
by smeared objects, thereby introducing a
level of uncertainty into spacetime.  As a
result of this uncertainty, a black hole
may become observationally indistinguishable
from a wormhole.  The wormhole, in turn,
may become observationally
indistinguishable from a gravastar.
It is also shown that the same
noncommutative-geometry background allows
the theoretical construction of thin-shell
wormholes from gravastars and may even
serve as a model for dark energy.

\end{document}